\documentclass{ptapap}
\author{Andrzej Pigulski}[IAUWr]
\author{Monika K.~Kami\'nska}[IOA]
\author{Krzysztof Kami\'nski}[IOA]
\author{Ernst Paunzen}[Brno]
\author{Jan Budaj}[TLo]
\author{Theodor Pribulla}[TLo]
\author{Pascal J.~Torres}[PUC]
\author{Ivanka Stateva}[BAS]
\author{Ewa Niemczura}[IAUWr]
\author{Marek Skarka}[Konk]
\author{Filiz Kahraman Ali\c{c}avu\c{s}}[COMU]
\author{Matej Seker\'a\v{s}}[TLo]
\author{Mathieu van der Swaelmen}[VUB]
\author{Martin Va\v{n}ko}[TLo]
\author{Leonardo Vanzi}[PUC]
\author{Ana Borisova}[BAS]
\author{Krzysztof He{\l}miniak}[TCAMK]
\author{Fahri Ali\c{c}avu\c{s}}[COMU]
\author{Wojciech Dimitrov}[IOA]
\author{Jakub Tokarek}[IOA]
\author{Aliz Derekas}[Szom,Konk]
\author{Daniela Fern\'andez}[PUC]
\author{Zoltan Garai}[TLo]
\author{Mirela Napetova}[BAS]
\author{Richard Kom\v{z}\'{\i}k}[TLo]
\author{Thibault Merle}[VUB]
\author{Milena Ratajczak}[IAUWr]
\author{Noel D.~Richardson \& Ritter Observing Team}[Ritt]
\author{Eiji Kambe}[Okay]
\author{Nobuharu Ukita}[Okay,GUAS]
\author{the BRITE Team}
\affil[IAUWr]{Instytut Astronomiczny, Uniwersytet Wroc{\l}awski, Wroc{\l}aw, Poland}
\affil[IOA]{Astronomical Observatory Institute, A.\,Mickiewicz University, Pozna\'n, Poland}
\affil[Brno]{Dept.~of Theoretical Physics and Astrophysics, Masaryk Univ., Brno, Czech Republic}
\affil[TLo]{Astronomical Institute, Slovak Academy of Sciences, Tatransk\'a Lomnica, Slovakia}
\affil[PUC]{Pontificia Universidad Catolica de Chile, Santiago, Chile}
\affil[BAS]{Institute of Astronomy with NAO, Bulgarian Academy of Sciences, Sofia, Bulgaria}
\affil[Konk]{Konkoly Observatory, Hungarian Academy of Sciences, Budapest, Hungary}
\affil[COMU]{\c{C}anakkale Onsekiz Mart University, Physics Department, \c{C}anakkale, Turkey}
\affil[VUB]{Inst.~d'Astronomie et d'Astrophysique, Univ.~Libre de Bruxelles, Brussels, Belgium}
\affil[TCAMK]{Department of Astrophysics, N.~Copernicus Astronomical Center, Toru\'n, Poland}
\affil[Szom]{ELTE E\"otv\"os Lor\'and University, Gothard Astrophysical Obs., Szombathely, Hungary}
\affil[Ritt]{Ritter Observatory, The University of Toledo, Toledo, Ohio, USA}
\affil[Okay]{Okayama Astrophysical Obs., National Astron.~Obs.~of Japan, Okayama, Japan}
\affil[GUAS]{The Graduate University for Advanced Studies, Tokyo, Japan}
\title{\boldmath{$\tau$}\,Ori and \boldmath{$\tau$}\,Lib:\\ Two new massive heartbeat binaries}

\begin{document}
\maketitle
\begin{abstract}
We report the discovery of two massive eccentric systems with BRITE data, $\tau$~Ori and $\tau$~Lib, showing heartbeat effects close to the periastron passage. $\tau$~Lib exhibits shallow eclipses that will soon vanish due to the apsidal motion in the system. In neither system, tidally excited oscillations were detected.
\end{abstract}
Following the discovery by \cite{2012ApJ...753...86T} of 17 stars in highly eccentric systems that show heartbeat effect during periastron passage and sometimes tidally excited oscillations (TEOs), there is a growing interest in the study of such systems. The first massive heartbeat system, $\iota$~Ori, was recently discovered by \cite{2017MNRAS.467.2494P}. We announce the discovery of two more relatively massive eccentric binaries showing heartbeat signals, $\tau$~Ori and $\tau$~Lib. 

$\tau$ Orionis (HD 34503, $V =$ 3.6~mag) is a visual quadruple system, in which the brightest component A is an evolved mid B-type star classified as B5 III. Variability of the radial velocity of $\tau$ Ori A was found over a century ago \citep{1907ApJ....25R..59F}, but up to now no orbital period was derived. In numerous studies, the star was used as an MK standard for spectral type B5\,III. It was not known to be variable in brightness. $\tau$~Librae (HD\,139365, $V =$ 3.6 mag), a member of the Sco-Cen association, is a double-lined spectroscopic binary classified as B3\,V\,+\,B5. The variability of the radial velocity of the star was found by \cite{1926PASP...38..132M}. Similarly to $\tau$~Ori, the star was not known as photometrically variable prior to the BRITE observations.  

\begin{figure}
\centering
\includegraphics[width=0.845\textwidth]{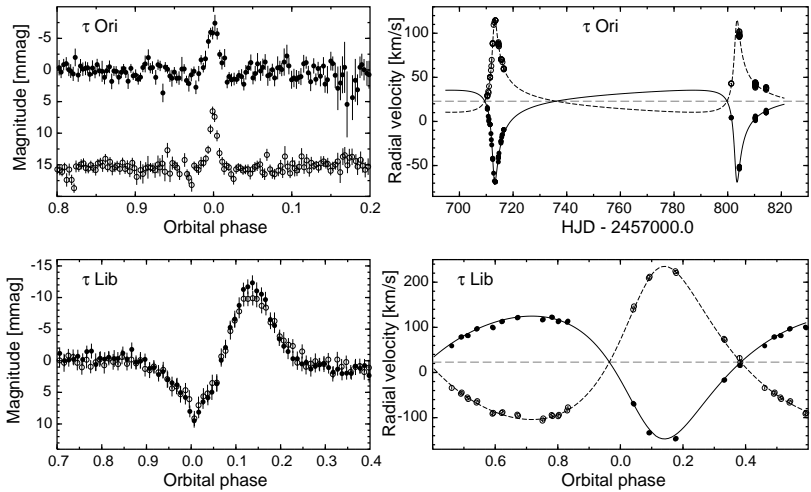}
\caption{Left panels: BRITE observations of $\tau$~Ori (top) and $\tau$~Lib (bottom) in the blue (filled circles) and red (open circles) bands phased with their  orbital periods: 90.3~d and 3.4501~d, respectively. Note the shallow eclipse at phase 0.0 for $\tau$~Lib. Right panels: Radial-velocity curves for the primary (filled circles) and the secondary (open circles) components of $\tau$~Ori (top) and $\tau$~Lib (bottom). The solid- and dashed-line curves stand for the preliminary orbital solution for the primary and secondary components, respectively.}
\label{fig:tau2}
\end{figure}

The two stars were observed by BRITE-Constellation \citep{2014PASP..126..573W,2016PASP..128l5001P} in 2013\,--\,2016 ($\tau$~Ori) and in 2015 ($\tau$~Lib). The photometry, obtained by means of the pipeline presented by \cite{2017A&A...605A..26P}, was subsequently corrected for instrumental effects and analyzed. For $\tau$~Ori, it revealed four short brightenings which we interpreted as a possible heartbeat signal in the vicinity of the periastron passage. The blue- and red-filter BRITE light curves of $\tau$ Ori phased with the derived orbital period of 90.3 d are shown in Fig.~\ref{fig:tau2}, top left panel. For $\tau$~Lib, the analysis revealed periodic variability with a period of 3.45~d, again showing a heartbeat shape, with a possibility of a very shallow ($\sim$3 mmag) eclipse (Fig.~\ref{fig:tau2}, bottom left panel). The eclipse was confirmed in the Solar Mass Ejection Imager (SMEI, \citealt{2004SoPh..225..177J}) observations made between 2003 and 2010, in which it is much deeper (10--14 mmag) than in 2015. This changing shape of the eclipse can be interpreted in terms of apsidal motion. 

The discovery of a heartbeat in $\tau$~Ori and $\tau$~Lib prompted us to organize a spectroscopic campaign. It resulted in over 260 spectra of $\tau$~Ori and 42 spectra of $\tau$~Lib. The radial-velocity curves (Fig.~\ref{fig:tau2}, right panels) show changes typical of eccentric systems. Both stars are double-lined spectroscopic binaries. $\tau$~Ori is a system consisting of two stars with practically the same mass (mass ratio $q =$ 0.997) in a highly eccentric orbit. The preliminary solution of the radial-velocity curve resulted in $P_{\rm orb} =$ 90.29 d, $e =$ 0.834, and $\omega =$ 156$^{\rm o}$. According to \cite{2012ApJ...753...86T}, the observed shape of the heartbeat corresponds to an inclination of about 30$^{\rm o}$, which implies component masses of about 6 M$_\odot$, consistent with the spectral classification. The preliminary spectroscopic parameters derived from the fit shown in Fig.~\ref{fig:tau2} are the following: $P_{\rm orb} =$ 3.4501~d, $e =$ 0.276, and $\omega =$ 155.4$^{\rm o}$. The inclination estimated at about 66$^{\rm o}$ allows to derive masses of 6.6 and 5.3 M$_\odot$. The difference between the BRITE and SMEI light curves can be explained by the change of the longitude of periastron. The estimated period of apsidal motion amounts to about 350 years. The eclipses will vanish in a few years. The star will become again eclipsing around 2050, when the secondary eclipse will start to be visible.

No tidally excited oscillations with amplitudes exceeding 0.35~mmag were found in the BRITE data of the two stars under consideration. A detailed analysis of the presented data will be published elsewhere.
\acknowledgements{
The study is based on data collected by the BRITE Constellation satellite mission, designed, built, launched, operated and supported by the Austrian Research Promotion Agency (FFG), the University of Vienna, the Technical University of Graz, the Canadian Space Agency (CSA), the University of Toronto Institute for Aerospace Studies (UTIAS), the Foundation for Polish Science \& Technology (FNiTP MNiSW), and National Science Centre (NCN). The operation of the Polish BRITE satellites is secured by a SPUB grant of the Polish Ministry of Science and Higher Education (MNiSW). The work is also based on observations obtained with the HERMES spectrograph, supported by the Fund for Scientific Research of Flanders (FWO), the Research Council of K.\,U.\,Leuven, the Fonds National de la Recherche Scientifique (F.R.S.-FNRS), Belgium, the Royal Observatory of Belgium, the Observatoire de Gen\`eve, Switzerland and the Th\"uringer Landessternwarte Tautenburg, Germany. We thank TUBITAK for a partial support in using RTT150 (Russian-Turkish 1.5-m telescope in Antalya) with project number 17ARTT150-1139. AP, EN, MR, and KH acknowledge the support from the NCN grants no.\,2016/21/B/ST9/01126, 2014/13/B/ST9/00902, 2015/16/S/ST9/00461, and 2016/21/ B/ST9/01613, respectively. This project has been supported by the Hungarian NKFI Grants K-115709 and K-119517 of the Hungarian National Research, Development and Innovation Office. Instrumentation development at the AI SAS was supported by the project ITMS No.\,26220120029. JB, TP, MV, and ZG acknowledge the support from the VEGA 2/0143/14 grant. ZG thanks for the support from the Slovak Central Observatory. IS, AB, and MN acknowledge the support of the Bulgarian NSF under grant DN 08-1/2016. MSka acknowledges the financial support of the Hungarian NKFIH Grant K-115709 and the postdoctoral fellowship programme of the Hungarian Academy of Sciences at the Konkoly Observatory as a host institution. AD was supported by the \'UNKP-17-4/II New National Excellence Program of the Ministry of Human Capacities. AD would like to thank the City of Szombathely for support under Agreement No.~67.177-21/2016.
}

\end{document}